\documentclass[twocolumn,a4paper,prb,aps,preprintnumbers,amsmath,amssymb,superscriptaddress]{revtex4-2}

\usepackage[utf8]{inputenc}
\usepackage{graphicx}
\usepackage{dcolumn}
\usepackage{bm}
\usepackage{hyperref}
\usepackage{times}
\usepackage{amsmath}
\usepackage{amsfonts}
\usepackage{amssymb}
\usepackage{textcomp}
\usepackage{gensymb}
\usepackage{color}
\usepackage[normalem]{ulem}
\definecolor{red}{rgb}{1,0,0}
\definecolor{blue}{rgb}{0,0,1}
\definecolor{darkred}{rgb}{0.6,0,0}
\definecolor{darkblue}{rgb}{0,0,.6}
\definecolor{darkgreen}{rgb}{0,0.5,0}
\definecolor{grey}{rgb}{0.5,0.5,0.5}

\begin{document}


\title{Energy dispersive X-ray spectroscopy of atomically thin semiconductors and heterostructures}

\author{Anna Rupp}
\def\LMU{Fakult\"at f\"ur Physik, Munich Quantum Center (MQC),
  and Center for NanoScience (CeNS),
  Ludwig-Maximilians-Universit\"at M\"unchen,
  Geschwister-Scholl-Platz 1, 80539 M\"unchen, Germany}
\affiliation{\LMU}

\author{Jonas G\"oser}
\affiliation{\LMU}

\author{Zhijie Li}
\affiliation{\LMU}

\author{Philipp Altpeter}
\affiliation{\LMU}

\author{Ismail Bilgin}
\affiliation{\LMU}

\author{Alexander H\"ogele}
\def\MCQST{Munich Center for Quantum Science and Technology (MCQST), Schellingtr. 4, 80799 M\"unchen, Germany}
\affiliation{\LMU}
\affiliation{\MCQST}

\date{\today}

\begin{abstract}
We report the implementation of energy dispersive X-ray spectroscopy for high-resolution inspection of layered semiconductors in the form of atomically thin transition metal dichalcogenides down to the monolayer limit. The technique is based on a scanning electron microscope equipped with a silicon drift detector for energy dispersive X-ray analysis. By optimizing operational parameters in numerical simulations and experiments, we achieve layer-resolving sensitivity for few-layer crystals down to the monolayer, and demonstrate elemental composition profiling in vertical and lateral heterobilayers of transition metal dichalcogenides. The technique can be straight-forwardly applied to other layered two-dimensional materials and van der Waals heterostructures, thus expanding the experimental toolbox for quantitative characterization of layer number, atomic composition, or alloy gradients for atomically thin materials and devices.
\end{abstract}

\maketitle

\section{Introduction}
The realm of two-dimensional materials with capacity for band engineering through elemental composition at the monolayer level and emergent hybridization phenomena in layered van der Waals heterostructures \cite{geim2013} represents a new paradigm in fundamental condensed matter research with applications in electronics \cite{radisavljevic2011,wang2012,kang2015} and optoelectronics \cite{lopez2013,britnell2013,bernardi2013,wu2015}. In semiconducting transition metal dichalcogenides (TMDs), the band gap can be tuned by layer number \cite{Mak2010,Splendiani2010} or the alloy composition of the respective crystal constituents \cite{komsa2012two,chen2013tunable,zhang2014two,tongay2014,xie2015}, which also allows to engineer the conduction band spin-orbit splitting \cite{Wang2015} and thus the valley polarization \cite{Liu2020} in monolayers or conduction and valence band offsets in respective heterostructures \cite{zi2019reversible}. The alloy composition can be adjusted in different TMD synthesis methods, including chemical vapor deposition (CVD) \cite{fu2015,apte2018,zhang2019recent} which, under optimized conditions, yields laterally extended monolayer crystals \cite{feng2014,li2014}, homobilayers and few-layer crystals \cite{gong2014band}, or lateral \cite{duan2014,li2015lateral} and vertical \cite{song2015} heterostructures. 

The resulting crystals often exhibit characteristic triangular shapes \cite{Zande2013,Najmaei2013}, allowing for simple identification of single-crystal monolayers with standard optical microscopy. More quantitative inspection of the layer number and composition in few-layer crystals can be performed by optical spectroscopy means including photoluminescence (PL) \cite{Mak2010,Splendiani2010,gong2014band,li2014,zhang2015,li2015lateral} and Raman mapping \cite{wang2013controlled,chen2014composition,liu2014,song2015,fu2015}. These techniques, bound in lateral resolution by the optical diffraction limit to a few hundred nanometers, are complemented by electron spectroscopy techniques such as X-ray photoelectron spectroscopy \cite{gong2014band,liu2014,feng2014,fu2015} or Auger electron spectroscopy \cite{tongay2014}. Energy dispersive X-ray (EDX) spectroscopy features a similarly high spatial resolution and additionally provides quantitative elemental analysis \cite{li2014}. Implemented in transmission electron microscopes (TEM), it has been successfully applied to two-dimensional materials \cite{shaw2014chemical,huang2014alloy,duan2014,song2015, bogaert2016, wang2017high} with the limitation of involved sample preparation methods as required for TEM experiments.

\begin{figure}[t]
\centering
\includegraphics[scale=1.01]{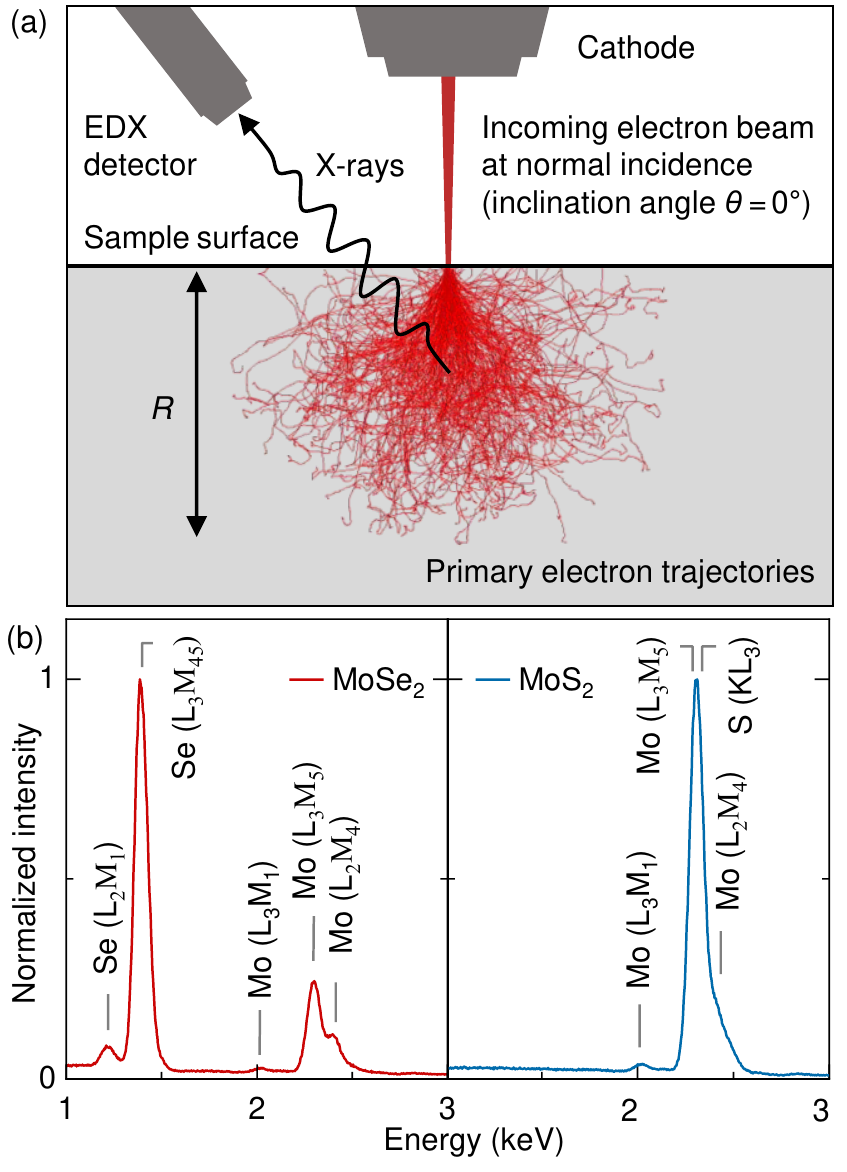}
\caption{(a) Illustration of energy dispersive X-ray spectroscopy in a scanning electron microscope (not to scale): upon excitation with primary electrons (red incoming beam at normal incidence for zero inclination angle $\theta$), X-rays (black arrow) reach the EDX silicon drift detector from the interaction volume (highlighted in red by primary electron trajectories) bound by the penetration range $R$. (b) EDX spectra of bulk transition metal dichalcogenide crystals MoSe$_2$ (left panel) and MoS$_2$ (right panel). For each spectrum, the intensity was normalized to the respective maximum; characteristic peaks of transition metal Molybdenum (Mo: L$_3$M$_1$, L$_3$M$_5$, and L$_2$M$_4$ at $2.020$, $2.293$ and $2.395$\:keV) and chalcogens Selenium (Se: L$_2$M$_1$ and L$_3$M$_{45}$ at $1.245$ and $1.379$\:keV) and Sulfur (S: KL$_3$ at $2.308$\:keV) are labelled explicitly.}
\label{bulk}
\end{figure}

In the following, we demonstrate how to adopt EDX analysis for TMD crystals in a standard scanning electron microscope (SEM). For layered TMD materials, sample preparation methods by exfoliation stamping \cite{Novoselov2005} or CVD synthesis on Si/SiO$_2$ and other substrates are well established without the need of modification for EDX spectroscopy. To date, however, the application of the SEM-EDX analysis to few-layer TMD crystals has been impeded by the small interaction volume bound by the monolayer thickness to below one nanometer \cite{sheng2017photoluminescence}. Performing optimization of the operation parameters in numerical simulations and calibration experiments, we establish EDX spectroscopy in SEM as a layer-resolving technique for TMD semiconductors, with sensitivity to alloy composition down to the monolayer limit. This feature, particularly beneficial for the characterization of CVD-grown TMD crystals with spatially varying alloy gradients, is confirmed by EDX profiling of a monolayer-thin lateral heterostructure.

\section{Experimental methods}

With access to efficient solid-state drift detectors in the 1960s, EDX spectroscopy became increasingly available for basic characterization of bulk materials with applications in material science, physics, chemistry, biology and medicine \cite{campbellEDX,chen2004alloy,yao2016,meiron2017,fonseca2018,fox2020}. In brief, EDX is based on X-ray detection in electron microscopes, as illustrated in Fig.~\ref{bulk}(a). The primary electron beam impinging on the sample undergoes different interactions upon propagation through the sample. The associated inelastic processes give rise to emission of secondary rays from the interaction volume, including element-specific X-rays relevant for EDX. X-ray radiation is generated when a vacancy created by primary electrons in the atomic inner-shells is refilled by outer-shell electrons. The energy of such transition is characteristic of the element and shells involved, and is dissipated either via Auger electrons (predominantly for light elements) or by X-ray radiation (more likely for heavier elements) which can be recorded with an EDX detector \citep{goldstein,reimer1984,reed}. The respective transitions give rise to peaks in the EDX spectra, classified by a capital letter (e.g. K, L, M) corresponding to the core level to which the de-excitation occurs with the subshell as a subscript (e.g. 1, 2, 3), followed by the letter and subscript of the original state. Using tables of element-specific transition energies and probabilities, EDX spectroscopy provides quantitative means for material composition analysis \citep{goldstein}.

To adopt EDX spectroscopy in our SEM (Zeiss, LEO DSM 982) equipped with an EDX detector (Oxford Instruments, X-Max$^N$ 50Standard with 50mm$^2$ detector area and an angle of 35° between the detector axis and the horizontally oriented sample) for a quantitative characterization of TMD crystals down to the monolayer limit, we first performed EDX signal calibration with MoSe$_2$ and MoS$_2$ bulk crystals. Bulk crystals were placed on thermal silicon oxide substrate ($285$\:nm SiO$_2$ on Si) and mounted in the SEM together with a copper tape above the substrate in close proximity to the sample to reduce beam drifts during the measurements. EDX spectra of bulk MoSe$_2$ and MoS$_2$, recorded with an acquisition time of $5$\:min for an aperture of $30$\:\textmu m and $10$\:keV electron beam energy, yielding 95\:pA sample current and a deadtime below 20\:$\%$ (acquisition software was used to correct for the deadtime and window transmission), are shown in the left and right panel in Fig.~\ref{bulk}(b) with characteristic peaks of Molybdenum (Mo), Selenium (Se) and Sulfur (S). The respective EDX peaks, on top 
of a weak but finite background of continuous X-ray bremsstrahlung, are proportional to the concentration of elements present in the probe volume of the sample. For element-selective analysis, the characteristic peaks were fitted by Gaussians to yield the total element-specific EDX signal as the sum over all peaks. This procedure resulted in a composition detection accuracy of $\pm\,1\:\%$ and $\pm\,2\:\%$ for bulk crystals of MoSe$_2$ and MoS$_2$ with spectrally distinct and overlapping peaks, respectively.

As opposed to TMD bulk crystals, EDX analysis of monolayers is much more challenging because of much smaller interaction volumes limited in one dimension to the few-atom layer. As highlighted in the schematic illustration in Fig.~\ref{bulk}(a), X-rays are generated along trajectories of primary electrons with energies sufficient for ionization of inner-shell atomic electrons. The corresponding electron trajectory range $R$ (in nm), as indicated in Fig.~\ref{bulk}(a), contains more than 95$\%$ of such trajectories and can be described by the Kanaya-Okayama equation \cite{kanaya}:  
\begin{equation}
    R= 27.6 \, E_0^{1.67} A/(Z^{0.89} \rho), 
    \label{range}
\end{equation}
where $E_0$ is the energy of the incoming electrons (in keV), $A$ the atomic mass, $Z$ the atomic number, and $\rho$ the material density (in g/cm$^3$). For TMD bulk crystals and an electron energy of $10$\:keV, this depth is about $650$\:nm which compares unfavorably with the TMD monolayer thickness below $1$\:nm. 

\begin{figure}[t]
\centering
\includegraphics[scale=1.01]{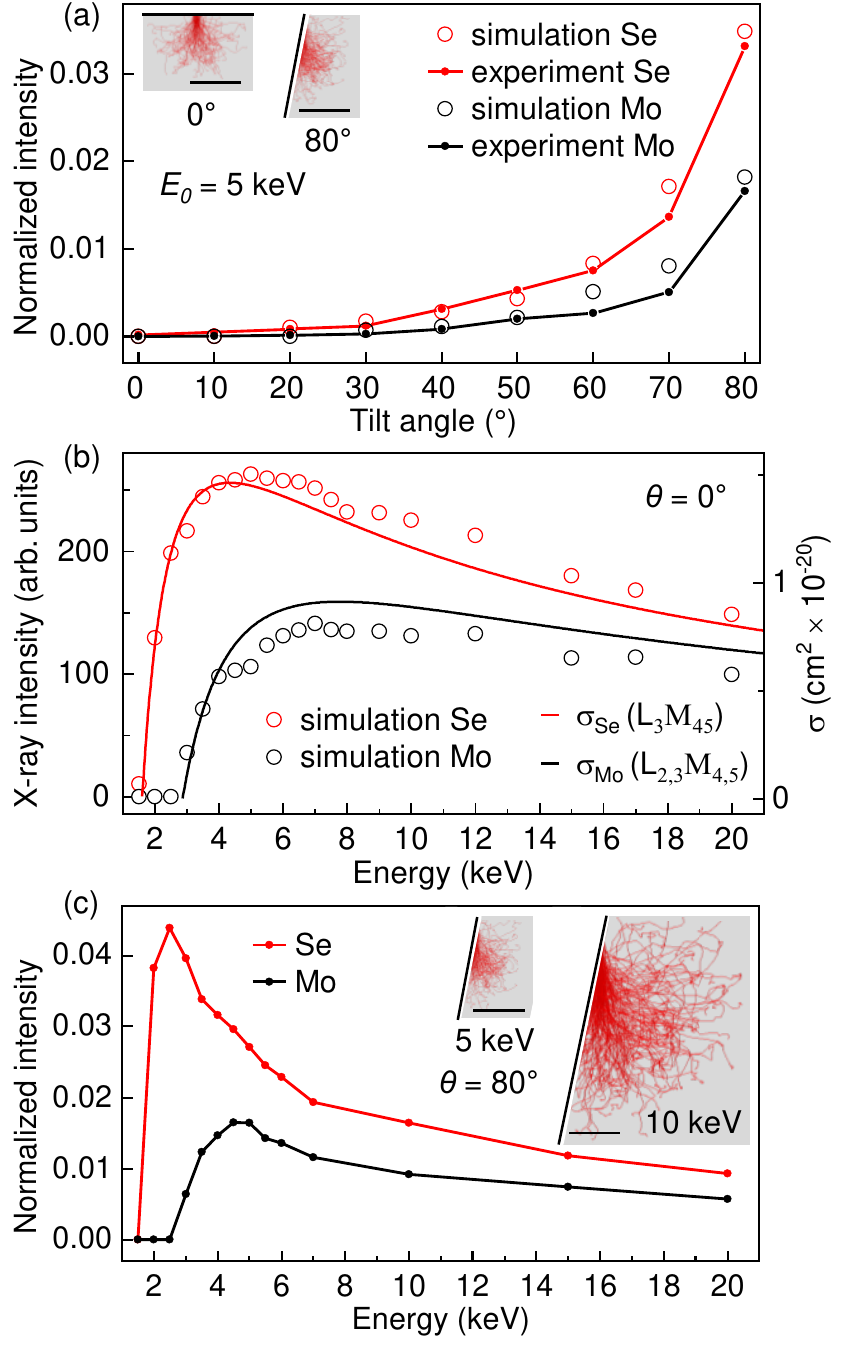}
\caption{(a) EDX intensity of Mo and Se peaks for monolayer MoSe$_2$ as a function of the sample tilt angle $\theta$ at electron beam energy of $5$\:keV. The insets show Monte Carlo simulations of the interaction volumes for $0\degree$ and $80\degree$ tilt. (b) Simulated EDX intensity (open circles) of Mo and Se peaks for monolayer MoSe$_2$ as a function of electron beam energy at zero tilt and calculated ionization cross section $\sigma$ (solid lines) for Se L$_3$M$_{45}$ and Mo L$_{2,3}$M$_{4,5}$ transitions. (c) EDX intensity of Mo and Se peaks for monolayer MoSe$_2$ as a function of electron beam energy at a tilt angle of $80\degree$. The insets show Monte Carlo simulations of the interaction volumes at $5$ and $10$\:keV beam energy for a tilt angle of $80\degree$. All scale bars are $300$\:nm, all data were recorded with an aperture of $30$\:\textmu m and $10$\:min acquisition time. The data in (a) and (c) were normalized to the KL$_3$ line of Oxygen in the underlying SiO$_2$ substrate.}
\label{angle}
\end{figure}

To increase the EDX signal for monolayers above the noise floor, we performed optimization of operation parameters both in Monte Carlo simulations and experiments. Numerical simulations of different operation conditions were carried out with a software package for quantitative X-ray microanalysis NIST DTSA-II \cite{NIST-software} for a MoSe$_2$ monolayer on 285\:nm SiO$_2$ on Si substrate. In our simulations, we identified two key factors for signal enhancement: increased interaction volume with the monolayer and optimization of the signal intensity according to non-linear ionization cross-section. The former can be achieved by tilting the sample away from the inclination angle of $\theta = 0 \degree$ at normal incidence, while the latter effect can accounted for by adjusting the energy of the incoming electrons, keeping in mind that lower energies would effectively increase the interaction volume with the monolayer at the sample surface by decreasing the penetration range according to Eq.~\ref{range}, yet maintaining sufficiently high energies for ionization. Guided by the simulations, we optimized both key parameters in experiments, with results shown in Fig.~\ref{angle}. 

The insets of Fig.~\ref{angle}(a) illustrate the interaction volumes obtained from simulations for tilt angles of $0\degree$ and $80\degree$ at $5$\:keV electron beam energy. Obviously, the interaction volume in a tilted geometry samples a larger area of the TMD monolayer on the sample surface. Consistently, the experimentally detected EDX intensity increases upon sample tilt from vanishingly small values at small angles by roughly two orders of magnitude for an inclination angle of $80\degree$, as evidenced for both Mo and Se elements by experimental and numerical results in Fig.~\ref{angle}(a). In addition to the increase of the trajectory length through the monolayer as the inverse cosine of the tilt angle, large tilt angles favor a reentrance of scattered electrons into the monolayer for successive X-ray generation. For the data in Fig.~\ref{angle}(a), the working distance between the cathode and the sample was optimized at each angle of inclination for maximum EDX signal. It is worth noting that the optimal working distance is specific to the configuration of the cathode and the EDX detector in the SEM and thus should be optimized consistently. For our SEM, a working distance of $7.5$\:mm proved optimal.

The results shown in Fig.~\ref{angle}(b) and (c) highlight the dependence of the X-ray intensity on the beam energy. For a tilt angle of $0\degree$, our simulations shown by open circles in Fig.~\ref{angle}(b) predict maxima in the X-ray signal as a function of the incoming electron beam energy: the element-specific EDX intensities exhibit onsets at the ionization energy $E_n$ of the respective element shell $n$ (at $1.436$\:keV for the Se L$_3$M$_{45}$ line, and $2.520$ and $2.625$\:keV for the Mo L$_3$M$_5$ and L$_2$M$_4$ lines \cite{bearden1967reevaluation}) and peak around twice to three times the ionization energy. The functional form of this behaviour is dictated by the ionization cross section, shown for both elements as solid lines in Fig.~\ref{angle}(b) and obtained (in cm$^2$) from the equation \cite{goldstein}:
\begin{equation}
\sigma_n = 6.51\cdot 10^{-20}\frac{z_n b_n}{E_0 E_n}\ln{\biggl(\frac{c_n E_0}{E_n}\biggr)},
\end{equation}
where $n$ is the shell number, $z_n$ the number of shell electrons, $E_0$ and $E_n$ (both in keV) are the beam and ionization energies of the shell, respectively, and $b_n$ and $c_n$ are effective Bethe parameters for a given element and shell \cite{powell1976cross}. The calculated cross-sections for Mo and Se transitions exhibit maxima around $7$ and $4$\:keV in very good agreement with the functional form of the simulated EDX intensities.

\begin{figure*}[t]
\centering
\includegraphics[scale=1.01]{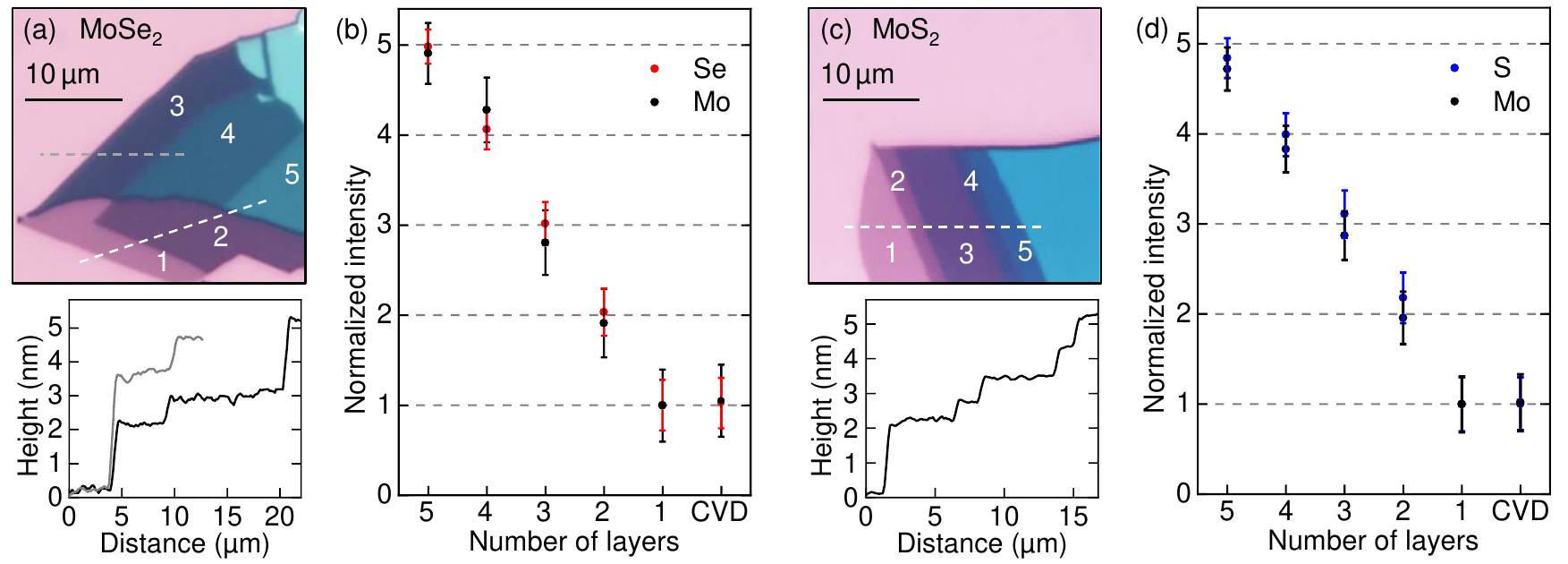}
\caption{(a) Top panel: Optical micrograph of a few-layer MoSe$_2$ crystal on SiO$_2$ with regions from one to five layer thickness. Bottom panel: Corresponding AFM topography scans along the dashed lines in the optical image. Note that the height of the first monolayer terrace is larger than the equidistant height steps for succeeding layers due to TMD-substrate interactions \cite{AFMoffset_Man}. (b) EDX intensity of Mo and Se peaks as a function of layer number down to the monolayer limit, shown together with a CVD-grown monolayer. All data were normalized to the peak intensity of the exfoliated monolayer. (c) and (d) Same as (a) and (b) but for MoS$_2$. All data were recorded with an aperture of $30$\:\textmu m and $10$\:min acquisition time at $75\degree$ tilt and $5$\:keV electron beam energy.}
\label{steps}
\end{figure*}

The data in Fig.~\ref{angle}(c) provide experimental proof for the enhancement in X-ray intensity anticipated from simulations. The signal was recorded on a MoSe$_2$ monolayer for a tilt angle of $80\degree$ and normalized to the K$\alpha$ line of oxygen in the underlying thermal oxide of the Si/SiO$_2$ substrate. Enhancement maxima were identified as a function of the electron beam energy in Fig.~\ref{angle}(c) around $3$ and $5$\:keV for Se and Mo signals, respectively. Although the energies of maximum enhancement are not identical with the maxima in the scattering cross-section of Fig.~\ref{angle}(b) due to the normalization procedure, the overall behavior of the enhancement in intensity is clearly confirmed. For simultaneous enhancement of Mo and Se signals, we chose an electron beam energy of $5$\:keV as an optimal tradeoff, which also favorably increases the interaction volume near the surface as shown by the insets of Fig.~\ref{angle}(c) for $5$ and $10$\:keV electron beam energies. We note that an electron beam energy around $5$\:keV was also found to be optimal for other TMD materials including MoS$_2$, WSe$_2$ and WS$_2$.

\section{Results}

With the optimized operation parameters for EDX analysis of TMD monolayers at hand, we demonstrate in the following layer-resolving performance of EDX spectroscopy on exfoliated few-layer MoSe$_2$ and MoS$_2$ crystals and CVD-grown monolayer. Further we demonstrate the lateral analysis of atomic composition in extended vertical heterostructures, homobilayers and lateral heterobilayers. Optical images of few-layer MoSe$_2$ and MoS$_2$ on Si/SiO$_2$ substrate are shown in the top panels of Fig.~\ref{steps}(a) and (c), respectively. The layer number was determined with PL in the monolayer limit and with AFM for multilayers. The AFM scans in the bottom panels of Fig.~\ref{steps}(a) and (c) consistently identify extended crystal terraces of one to five layers. 

\begin{figure*}[t]
\centering
\includegraphics[scale=1.01]{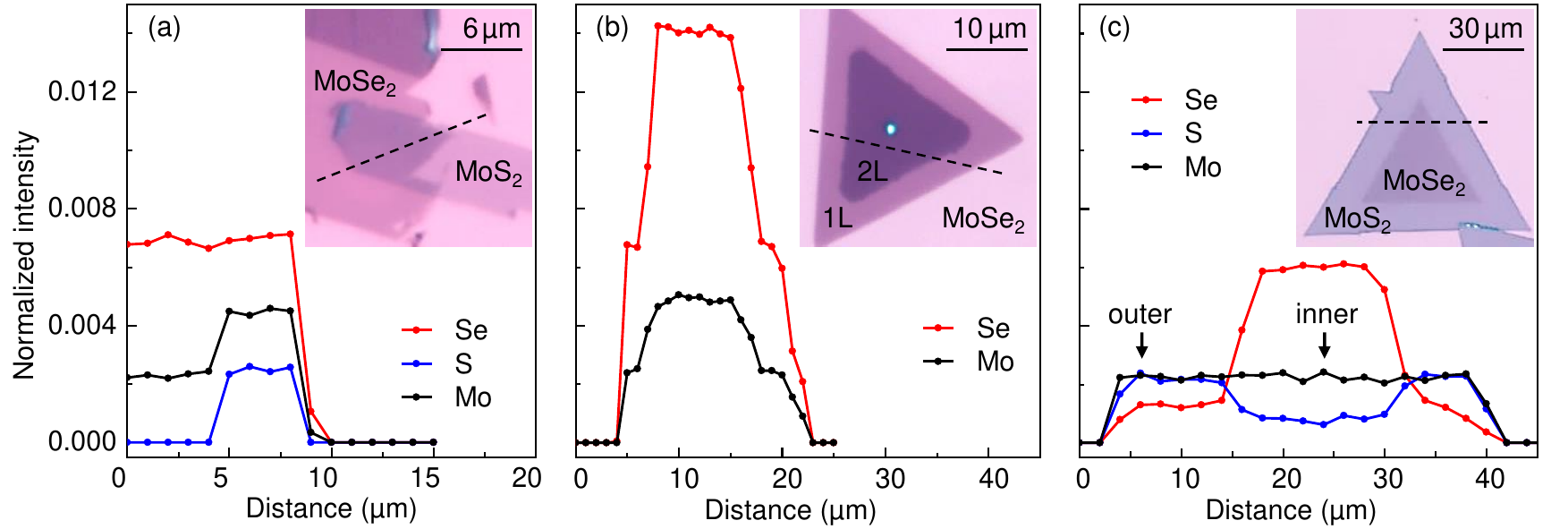}
\caption{(a) EDX profile of an exfoliation-assembled MoSe$_2$-MoS$_2$ vertical heterostructure on SiO$_2$ along the dashed line in the optical micrograph (inset). (b) and (c) Same for CVD-grown MoSe$_2$ homobilayer and MoSe$_2$-MoS$_2$ lateral heterobilayer. The arrows in (c) indicate representative positions where the alloy concentration was determined for the inner and outer regions of the MoSe$_2$-MoS$_2$ heterobilayer. All data were normalized to the KL$_3$ line of Oxygen in the underlying SiO$_2$ substrate.}
\label{HBL}
\end{figure*}

At each terrace, EDX signal acquisition was performed for $10$\:min with an electron beam energy of $5$\:keV (with 85\:pA sample current through a 30\:\textmu m aperture and a deadtime below 5\:$\%$) for a sample tilt angle of $75\degree$ with results shown in Fig.~\ref{steps}(b) and (d) for exfoliated few-layer crystals with consecutively decreasing number of layers. Additionally, we recorded single-crystal CVD-grown monolayers of MoSe$_2$ and MoS$_2$ on complementary samples (rightmost data points). All data were normalized to the respective EDX intensity of exfoliated monolayers with a detection accuracy of $\pm 5\:\%$ and $\pm 8\:\%$ for MoSe$_2$ and MoS$_2$ monolayers, respectively. The observation of equidistant steps in the EDX intensity as a function of the number of layers unambiguously confirms the single-layer sensitivity of our measurements as well as identical EDX signals (within error bars) for exfoliated and CVD-grown monolayers.

This quantitative calibration of material-specific EDX signals down to monolayers of exfoliated and CVD-synthesized TMDs provides means for layer number and composition analysis in laterally extended homo- and heterobilayer crystals. To this end, we fabricated on Si/SiO$_2$ substrates a vertically stacked MoSe$_2$-MoS$_2$ heterobilayer by standard exfoliation stamping of individual monolayers as well as a vertical MoSe$_2$ homobilayer and a lateral MoSe$_2$-MoS$_2$ heterobilayer by CVD synthesis. The optical micrographs of each sample are shown as insets in Fig.~\ref{HBL}(a), (b) and (c). The respective EDX profiles were recorded along the dashed lines in the insets as consecutive spectra upon lateral displacement of the electron beam with respect to the sample. For all raster-step EDX measurements, the inclination angle was reduced to $45\degree$ for higher spatial resolution below $300$~nm as estimated from numerical simulations for an electron beam energy of $5$\:keV. With this beam energy, EDX data were recorded in discrete steps with an acquisition time of $30$\:min at each spot.

The optical micrograph in Fig.~\ref{HBL}(a) shows the vertical  MoSe$_2$-MoS$_2$ heterobilayer region and indicates by the dashed line the lateral trajectory of the EDX profile over $15$~\textmu m recorded in consecutive steps of $1$~\textmu m from the lower left to the upper right point of the line. It starts out with Mo and Se intensities characteristic of monolayer MoSe$_2$ and jumps after $5$~\textmu m in just one lateral step by the excess contributions of Mo and S of monolayer MoS$_2$. After additional four lateral steps, the EDX signal drops to zero within one step away from the heterostructure. According levels of discrete changes in the EDX intensity profile were detected for the CVD-grown MoSe$_2$ homobilayer of Fig.~\ref{HBL}(b) with a total distance of $25$~\textmu m in $0.8$~\textmu m steps. The transverse passage of the homobilayer nearly orthogonal to the left triangle edge resulted in sharp jumps of the detected EDX intensity, doubling the characteristic signals of Mo and Se in two consecutive steps and thus unambiguously identifying the transition from monolayer to bilayer. As expected, the EDX profile shows the reverse behavior upon further transition away from the flake with a simultaneous drop of characteristic Mo and Se signals to zero at the bare substrate. It is worth noting the vanishing contamination of the CVD-grown terraces and the underlying substrate by other elements.  

Finally, we demonstrate that optimized EDX spectroscopy is powerful in quantifying alloyed layer composition. To this end, we inspected  a CVD-grown lateral MoSe$_2$-MoS$_2$ heterobilayer shown in the optical micrograph of Fig.~\ref{HBL}(c) with the inner MoSe$_2$ monolayer triangle grown first (dark triangle) and the outer lateral MoS$_2$ boundary (lighter regions) added in a subsequent growth step. The corresponding EDX profile was performed over $45$~\textmu m in steps of $2$~\textmu m along the dashed line. Contrary to well-defined boundaries between MoS$_2$ and MoSe$_2$ regions expected from sequential growth, the EDX profile reveals cross-contamination of the adjacent regions by S and Se chalcogens. For points in the outer and inner regions of the lateral heterobilayer indicated by arrows in Fig.~\ref{HBL}(c), we determined composition fraction $x$ of the MoSe$_{2x}$S$_{2(1-x)}$ alloy as $0.15 \pm 0.05$ and $0.82 \pm 0.05$, respectively. At the boundary between the inner and outer regions, the EDX profile clearly reflects a gradient in the S and Se concentrations in the presence of a constant Mo concentration. Given the spatial resolution of $300$~nm and a step size of $2$~\textmu m, EDX profiling thus detects unambiguously varying alloy concentration not obvious in the optical micrograph with sharp delimiting boundaries between the inner triangle and the outer monolayer region with conformal geometry. Similar observations were made on a lateral CVD-grown WSe$_2$-WS$_2$ heterobilayer (data not shown), confirming cross-contamination of the chalcogen atoms in the growth process \cite{gong2014band, bogaert2016}. These results highlight the generic analytic power of layer- and element-sensitive EDX profiling of TMD heterostructures down to the monolayer limit.

\section{Conclusions}
In conclusion, we reported optimized implementation of EDX spectroscopy integrated in a SEM for elemental profiling of semiconducting TMD crystals down to the monolayer limit. The layer-resolving sensitivity was achieved by optimizing operational parameters in both simulations and experiments. Based on quantitative calibration experiments of element-specific EDX intensities on bulk and few-layer TMD crystals, we demonstrated the applicability of the technique to layer number, elemental composition and alloy gradient detection by mapping out EDX profiles of vertical and lateral TMD heterostructures synthesized by CVD or fabricated by exfoliation stacking. Since EDX spectroscopy is not limited to the specific materials used in our study, we anticipate that SEM-based EDX analysis of varying element and alloy compositions in layered crystals will become a valuable characterization method for the entire class of two-dimensional materials and their van der Waals heterostructures.   

\section{Acknowledgements}
This research was funded by the European Research Council (ERC) under the Grant Agreement No.~772195 as well as the Deutsche Forschungsgemeinschaft (DFG, German Research Foundation) within the Priority Programme SPP~2244 2DMP and the Germany's Excellence Strategy EXC-2111-390814868. A.\,R. acknowledges funding by the Munich Quantum Valley doctoral fellowship program within the Bavarian initiative "Hightech Agenda Bayern Plus". Z.\,Li. was supported by the China Scholarship Council (CSC), No. 201808140196. I.\,B. acknowledges support from the Alexander von Humboldt Foundation, and A.\,H. from the Center for NanoScience (CeNS) and the LMUinnovativ project Functional Nanosystems (FuNS).

\end{document}